\documentstyle[preprint,aps]{revtex}
\begin{document}
\draft
\title{Inductive Probing of the Integer Quantum Hall Effect}

\author{E. Yahel $^1$, D. Orgad$^2$, A. Palevski$^1$, 
 H. Shtrikman$^2$ }

\address{$^1$School of Physics and Astronomy, Raymond and Beverly Sackler 
Faculty of Exact Sciences,\\
Tel Aviv University, Tel Aviv 69978, Israel\\ 
$^2$Department of Condensed Matter, The Weizmann Institute of Science,
Rehovot 76100, Israel.\\}

\date{\today}
\maketitle

\begin{abstract}
We investigated the Integer Quantum Hall Effect (IQHE) using an
inductive method. The following conclusions can be derived from
our study: 
(i) when the Fermi energy is located between Landau levels 
the only extended states at the Fermi energy are
located at the physical edges of the sample. 
(ii) the extended states located at the
bulk of the sample below the Fermi energy are capable of carrying 
a substantial amount of 
Hall current, but cannot screen an external electrostatic potential.

\end{abstract}


\narrowtext

Since the discovery of the Integer Quantum Hall Effect (IQHE) \cite{Klit}, the
role of bulk \cite{Ando,Avron,Thoul1,Thoul2} versus 
edge \cite{Halp,Mac,Butt,Shi} states has been discussed theoretically.
The results of many experiments
\cite{Ebert,Zheng,Wees,Alph,McEuen,Tak} addressing this issue seem to 
favor the edge picture over the bulk one.
However, recent experimental studies  
\cite{Foun,Haren,Udi} revived this controversial question by
giving evidences supporting the bulk picture.
%In these studies it has been shown that 
%in the dissipative regime, i.e. when the Fermi energy crosses a Landau level,
%the variation of the electrostatic potential
%across the sample is linear (the electric field is uniform). 
%At the Hall plateaus when the Fermi energy is located between consecutive 
%Landau levels, 
In these studies it has been shown that the electrostatic potential varies 
in the bulk of the sample. It implied the existence of Hall current 
carried by the bulk states. 
%Capacitance measurements \cite{Tak} of a 2DEG in the IQHE regime indicate 
%that in the nondissipative regime, the extended electronic states at the 
%Fermi energy occupy only a small fraction of the sample area along its edge.
%This measurement however, did not produce information about the ability of 
%the bulk states to carry current. 

The magnetic coupling between a SQUID magnetometer and a 2DEG has been 
suggested for studies of current distributions \cite{Mendez}. However, 
this method is extremely difficult to realize experimentally since 
it requires the critical field $H_{c}$ of the SQUID to be higher than 
the typical magnetic fields used in IQHE experiments.
Another variation of inductive coupling has been employed in a recent
experiment \cite{Hall} where an external solenoid was used in order to induce 
azimuthal electric field in 2DEG samples patterned in a Corbino geometry.
Although the authors observed well-defined Hall plateaus, 
they did not provide any information about the
spatial distribution of the extended states at the Fermi energy.

In order to address the questions concerning the role of edge versus bulk
states in the IQHE we employed an inductive coupling, different from those
mentioned above. Our method utilizes a pick-up coil in order to measure  
time-dependent magnetic fields induced by alternating currents in the sample.
Although the sensitivity limitations of this method do not allow for a 
precise determination of the current's spatial distribution, a
quantitative analysis of our data allows us to reaffirm the following 
important statements: 
(i) in the plateaus of the IQHE, the extended states at the Fermi energy are 
located at the edges of the sample. 
(ii) in this regime the bulk states at the Fermi energy are localized. 
However, the bulk states, at the Landau levels
below the Fermi energy, may carry a substantial amount of the Hall current. 
The contribution of these bulk
states to the Hall current depends on the details of the electrostatic
potential. The latter is strongly influenced by the geometry of the sample
and by the attached contacts.

The 2DEG samples used in this study were fabricated from 
${\rm GaAs}_{x}/{\rm Al}_{1-x}{\rm GaAs}$ heterostructures. 
The electron carrier concentration and the mobility of the samples were
$n=2.1 \times 10^{11}$ ${\rm cm}^{-2}$ and $\mu=6.4 \times 10^5$ 
${\rm cm}^2$/Vs at 1.4 K respectively.
Rectangular shaped samples with typical dimensions of $10
\times 5$ ${\rm mm}^2$ were cleaved from the wafer and Ohmic Au/Ge/Ni 
contacts were alloyed at opposite sides.
A 3000 turns pick-up coil was placed 0.4 mm above
the sample's physical edge.
The effective area of the pick-up coil was $5 \times 5$ ${\rm mm}^2$. 
A schematic view of the geometrical setup is shown
in Fig. 1.

An alternating current at frequency $\omega$, driven through the sample, 
produced an electro-motive force at the same frequency in the pick-up coil 
circuit.
A grounded metallic shield made of brass foil was used to screen any direct 
electrostatic coupling between the pick-up coil and the sample. 

The voltage which develops across the pick-up coil depends on the 
distribution of the currents in the bar and on geometrical factors of the 
setup. Although the value of the pick-up voltage can be estimated 
theoretically \cite{Mendez} for any given distribution of the current, 
we have performed an experimental calibration of the response of our 
pick-up coil. 
We have found that for homogeneous current distribution, at
frequency of 6.4 KHz, the voltage response of the pick-up coil was 
25 nV/$\mu$A. In order to demonstrate the sensitivity of the pick-up coil to 
changes in the current distribution, we have deposited a 1000 ${\rm \AA}$ 
thick and 500 $\mu$m wide Au film along the periphery of a sample having the 
same geometry. In this case, the pick-up response increased 
to 45 nV/$\mu$A at the same frequency. Although this
calibration gives smaller pick-up response values than those calculated
theoretically, the relative change of the signals between a uniform and edge 
distributions of currents is consistent with the theoretical estimate. 
We believe that the discrepancy between the theoretical and
experimental absolute values of the pick-up response is due to partial 
screening of the inductive coupling by eddy currents in the metallic 
shield.
These currents were found to be sensitive to the conductivity of the shield 
and varied with temperature.
The calibration values mentioned above are given 
for low temperatures where the pick-up response was found to be 
temperature independent.
Since the distance of the shield from the sample is relatively large 
($\sim 400\mu$m) and because the 
dielectric constant of the media is an order of magnitude smaller 
then that of GaAs we do not expect the shield to significantly alter the
potential distribution in the sample. 

A standard four probe measurement of the IQHE in our Hall bar samples, 
resulted in the experimental curve for the longitudinal resistivity
$\rho_{xx}$ 
shown in the inset of Fig. 2. 
Since the lowest temperature of our experimental setup was 
1.4 K and the highest magnetic field was 5.5 T, only the plateaus with 
$\nu=2, 4$ showed experimentally zero values of the longitudinal 
resistivity $\rho_{xx}$.

In the first part of our investigation a metallic gate has been deposited 
on the bottom surface of the sample (back gate), 250 $\mu$m from the 2DEG. 
We have applied an alternating voltage $V_g$ between the back gate and the 
2DEG and monitored the signal in the pick-up coil $V_{pc}$ as the magnetic 
field $H$ was swept in the range between -5 to 5 Tesla. The results of this 
measurement are shown in Fig. 2. The amplitude and frequency of the 
applied gate voltage were varied in the range of 0.05-0.5 V and 0.2-30 KHz, 
respectively. 
%Within this range, the signal was found to depend linearly on 
%these parameters. At voltages exceeding 1V, deviation from linearity was 
%observed and the dependence of the signal on the applied voltage was weaker.
%A possible source for this nonlinearity could be the onset of the breakdown 
%of the IQHE.
%Such a breakdown is expected to result in a current distribution which is
%extended into the bulk. This in turn, will decrease the signal measured 
%by the pick-up coil.

The peaks in  $V_{pc}$ are clearly observed for values of $H$ for which 
the longitudinal resistivity vanishes. According to our calibration, the 
values of the peaks correspond to a current 
$V_g \nu e^2/h $ flowing around the periphery of the sample, 
where $\nu$ is the number of occupied Landau levels
below the Fermi energy.

At first, this result seems to be surprising since for the estimated values
of the capacitance of our samples such values of $V_g$  cannot 
produce or modulate a Hall current of the observed magnitude. Indeed,
there is no signal at the pick-up coil at the entire range of the magnetic 
field besides the regions corresponding to Hall plateaus. The resolution to 
this apparent ``mystery'' becomes clear when one assumes that at the Hall 
plateaus the entire bulk of the sample becomes an insulator, while the edges 
are conducting. 
In such a situation, the electric potential of the sample 
should approach the value of $V_g$ as the distance from the edge becomes
larger than the distance to the back gate. Applying $V_g$ under such
conditions is equivalent to the application of Hall voltage to a Corbino 
geometry sample. It results in a Hall current of the observed magnitude, 
circulating along the sample's boundaries.
Since the direction of the current should be reversed when the polarity
of the magnetic field is changed, the pick-up signal should also reverse its 
sign under such an operation as indeed one finds by inspecting Fig. 2. 

The existence of extended states in the bulk of the sample at the Fermi
energy is equivalent to introducing extra edges to the sample. 
Furthermore, it would
increase the distance from the edge at which the electric potential attains
its maximal value $V_{g}$. In addition, if the extended states below the
Fermi energy were able to partially screen the external voltage, the Hall
voltage developed in the sample, would have been smaller then $V_{g}$.
All of these effects would tend to diminish the signal measured by the 
pick-up coil.
The measured signal, however, is the largest possible since the Hall voltage 
cannot exceed $V_g$ and the current cannot flow any closer to the pick-up 
coil then along the sample's edge. This observation leads us to 
two important conclusions: 
{\it (i) at the Hall plateaus the only extended states at the Fermi energy 
are located along the sample's edges. (ii) the extended states below the
Fermi energy, though capable of carrying Hall current,} as will be shown
later, {\it cannot screen the external electric field}. This is the first
direct observation of this property which is implicit in the bare existence
of the IQHE.
Our experimental resolution provides us with an upper bound of 0.5mm (10\% of
the sample's width) to the distance from the edge in which the current flows.

Within the measuring range of applied voltages and frequencies, the signal 
was found to depend linearly on 
these parameters. However, at voltages exceeding 1V, deviation from 
linearity was observed and the dependence of the signal on the applied 
voltage was weaker (not shown).
A possible source for this nonlinearity could be the onset of the breakdown 
of the IQHE.
Such a breakdown is expected to result in a current distribution which is
extended into the bulk. This in turn, decreases the signal measured by the
pick-up coil.

Although the experiment described above indicates that the Hall current flows
in the vicinity of the edge, it should not be concluded that such a 
non-uniform distribution between the bulk and the edges is an inherent 
property of the IQHE.
On the contrary, it is the proximity of the back gate to the 2DEG (their
separation is much smaller then the dimensions of the sample) that causes the
electrostatic potential to be flat far from the edges and to change by
$V_g$ in the vicinity of the sample's edges.
Therefore we cannot resolve questions concerning the contribution of bulk 
states to the current based on this experiment.

In order to address this issue we fabricated two samples in which the 
back gate was replaced by additional Ohmic contacts which were alloyed in the 
interior of the 2DEG. 
In the first sample, the inner contact occupied almost its entire area, 
thus defining a strip of 2DEG along the edges.
Such a geometry is usually referred to as Corbino geometry.
The experimental setup is presented in Fig. 3.
We applied a source voltage and measured the current flowing in the circuit
and the voltage drop across the shunt resistor.
An alternating voltage drop $V_{r}$, which developed between the inner 
contact and the Ohmic contact located at the sample's edge, resulted in 
a pick-up signal that corresponded to a Hall current of the same value as in 
the previously described experiment, namely, $I=V_r \nu e^2/h$. Fig. 3. shows 
the pick-up coil response and the Hall current calculated using
\begin{equation}
 I_{H} = \frac{V_{r}-R_{xx} I_{r}}{R_{xy}} = \frac{V_{source}-V_{shunt}(1+
           \frac{R_{0}+R_{xx}}{R_{shunt}})}{R_{xy}}\;\; ,
\end{equation}
where $I_{r}$ is the dissipative current that flows between the inner contact 
and the edge. 
This current vanishes for Corbino geometry samples at the IQHE plateaus and
the expected value of the Hall current is $I_{H}=V_{r}/R_{xy}$.
At these regions $V_{r}$ equals $V_{source}$ (c.f. Fig. 3).
In the dissipative regimes, namely in between plateaus and at small values of
magnetic fields, the Hall current 
is practically independent of $R_{xx}$ as long as the latter is much smaller
then $R_0$. Since $R_{0}=0.5$M$\Omega$ and $R_{xx}$ is of the order of
$\rho_{xx}$, measured in Hall bar geometry, we expect this inequality to hold.
Accordingly, we also expect $V_{r}$ in these regions to be much smaller then 
$V_{source}$, thus resulting in a smaller signal in the pick-up coil.
For $R_{xy}$, we use the values obtained from the four probe measurement. 
The good agreement between the pick-up signal and 
$I_{H}$ given by Eq. (1) indicates that indeed the pick-up coil measures
the circulating current in the sample. One should note that the latter 
consists of a constant diamagnetic current and a time-dependent Hall current 
induced by $V_{r}$. The pick-up coil is sensitive of course only to the 
second component.
This measurement also provided us with an additional calibration of the 
pick-up response, which was consistent within few percents with the previous 
calibration procedure.

A second sample with two inner contacts having dimensions of 100$\mu$m
$\times 100\mu$m (see inset of Fig. 4) has been measured using the same 
technique. The pick-up signal versus the magnetic field when an 
alternating voltage $V_{source}$ was applied to the device is shown in Fig. 4.
The pick-up coil signal at integer filling factors in this case dropped
significantly relative to the value measured for the sample shown in Fig. 3.
Since the total Hall current in this configuration should be the same in both
cases, the only possible explanation is a spatial redistribution of the
current.
Moreover, the signal detected by the pick-up coil varies considerably for
different realizations of the circuit as depicted in Fig. 4.
For configuration c the signal is smaller by an order of magnitude relative
to the signal measured in the case of the sample with the large inner
contact. This undoubtedly proves that Hall current is carried by bulk states.
As far as we know, this is the first direct experimental evidence for bulk
current in the IQHE. 
Although the spatial resolution of our technique does not allow for a
precise determination of the current distribution, we can set an upper 
bound for the edge current.
Under the assumption that the only contribution to the pick-up signal is due
to edge states, we find that, at most, 10\% of the total Hall current is 
carried by the edge. However, it is more reasonable to conclude that the
actual fraction carried by the edge is much smaller 
(if not zero \cite {Thoul3})
and the entire signal in the pick-up coil is due to bulk current.
The results for the different configurations of contact connections indicate 
that the current distribution in the sample depends on the details of the 
electrostatic potential, which can be strongly influenced by the 
geometry of contacts, the presence of gates, etc.

We would like to emphasize that our conclusions about the role of edge versus
bulk states apply to Corbino geometry samples and further investigation
addressing this problem for Hall bar geometry is required. The main difference
between the two geometries, at the IQHE regime, is the necessity to inject
external current from the Ohmic contact into the 2DEG for the Hall bar 
geometry. 
The current injection mechanism could significantly enhance \cite {vanson} 
the role of the edge currents. 

We have benefited from useful discussions and help from Y. Kornblit, Y. Imry, 
U. Sivan. M. Heiblum O.Entin-Wholman and Y. Berk. The research was 
partially supported by the Israeli Academy of Sciences and Humanities.

\figure{Fig. 1. The experimental setup. Lateral (a) and top (b) views of the 
                2DEG and the pick-up coil.}

\figure{Fig. 2. Inductive voltage in the pick-up coil for a
                back gate voltage $V_{g}=0.5$V at a frequency of 6.4KHz. 
                The left axis depicts the number of filled Landau levels
                needed to produce the same signal, assuming that the current 
                flows within a distance of 500$\mu$m from the edge.
                $V_{pc}$ clearly shows well resolved picks in the middle 
                of Hall plateaus.
                Plateaus with $\nu=2,4$ are already saturated whereas 
                $\nu=6$ still has nonzero longitudinal resistance.
                The insets show a schematic view of the sample  
                and the longitudinal resistivity measured on
                the same wafer.}

\figure{Fig. 3. Solid line - Hall current in a sample having a Corbino
                geometry as deduced from the pick-up signal. The source
                voltage is 0.5V at a frequency of 26KHz.
                Dotted line - Hall current calculated according to Eq. (1).
                The inset shows a schematic view of the experimental
                measuring circuit.}

\figure{Fig. 4. Pick-up coil signal for various contacts configurations as
                shown in the inset. The source voltage is 0.5V at a frequency
                of 26KHz. 
                a) Solid line - voltage applied to both point contacts. 
                b) Dashed line - one contact left floating.
                c) Dotted line - one contact grounded. }

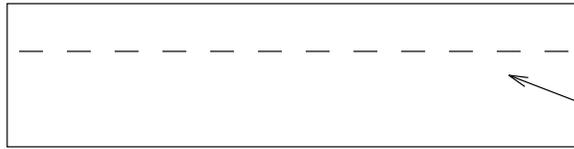
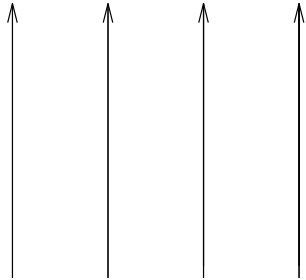
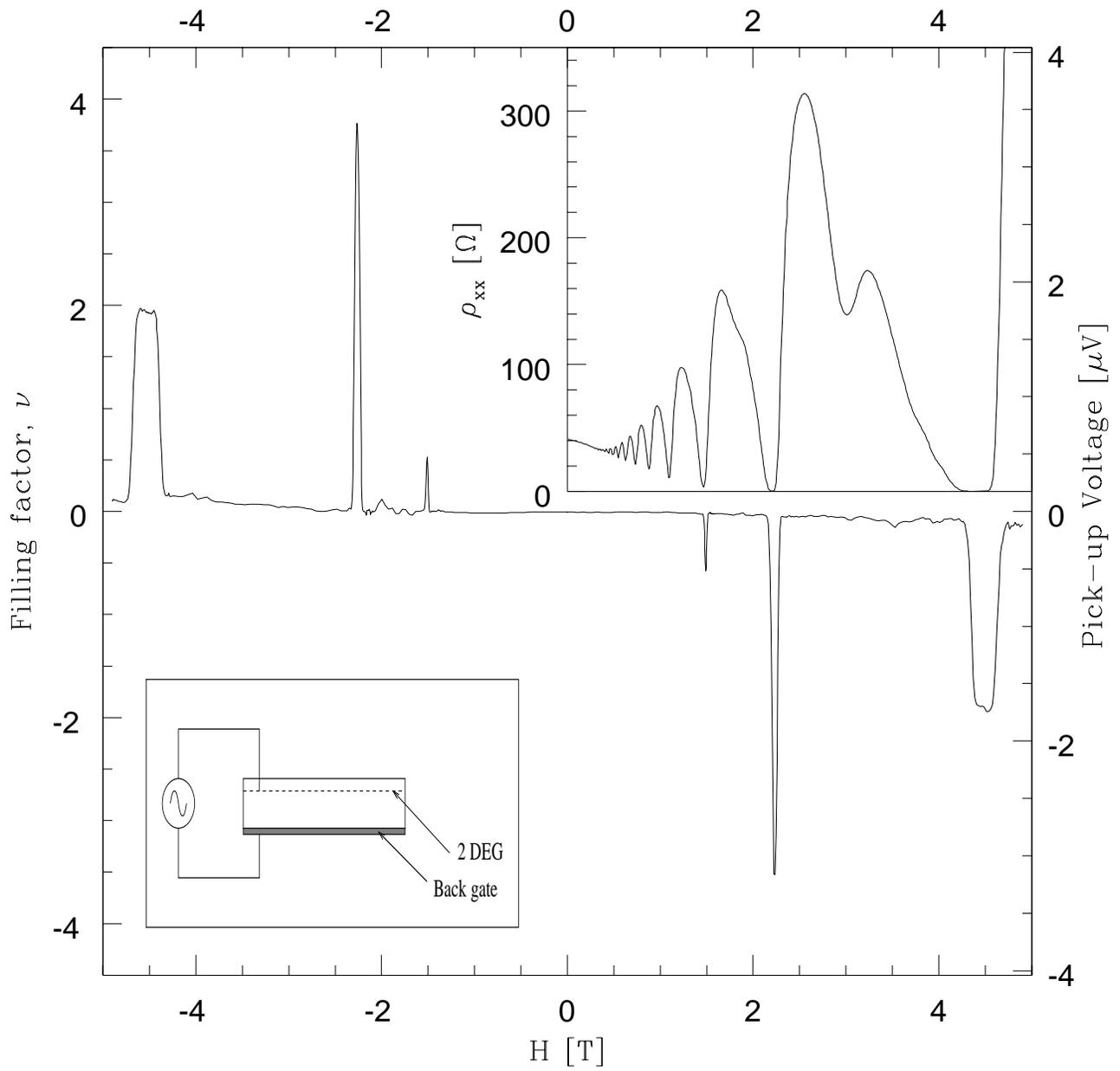
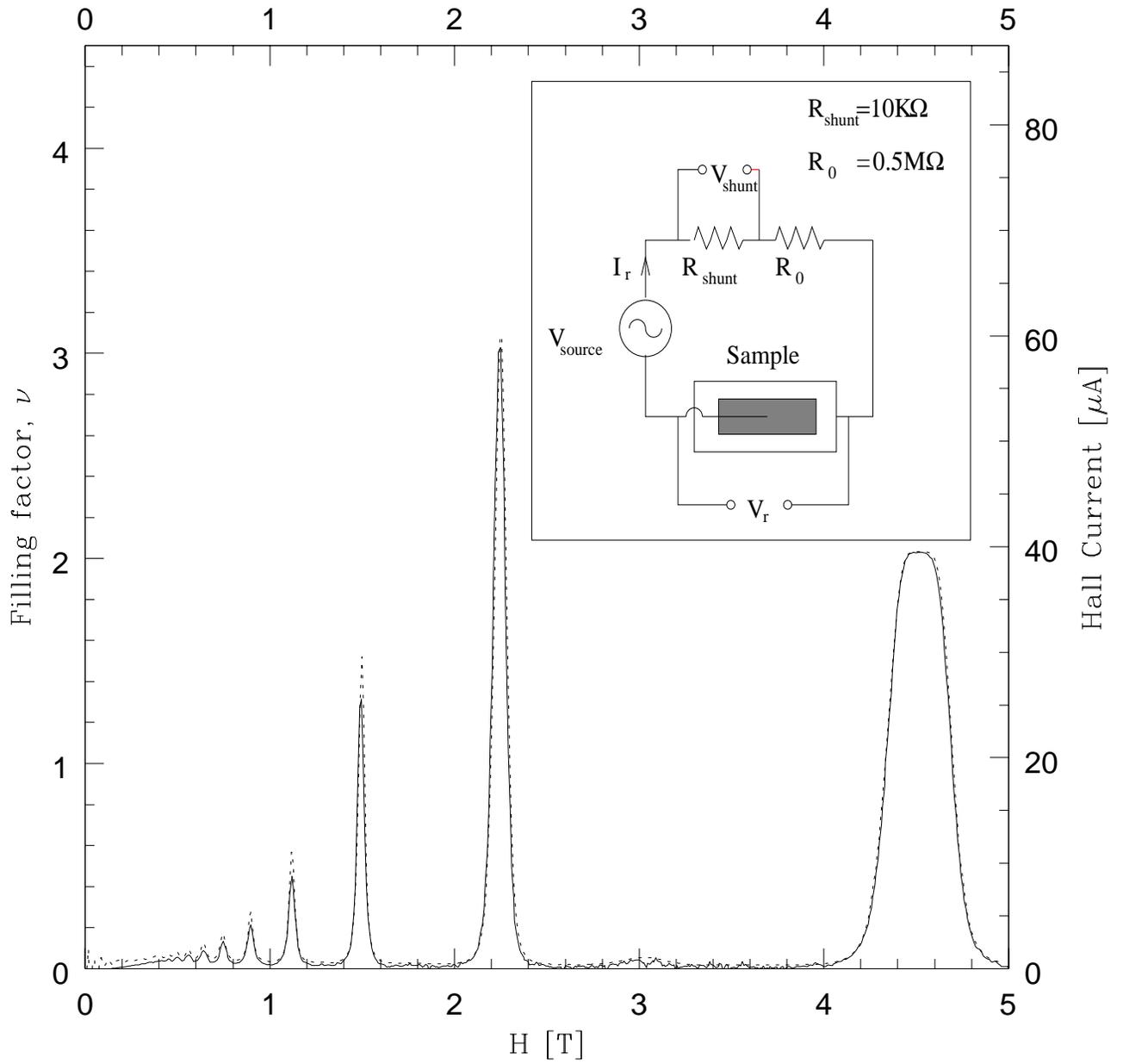
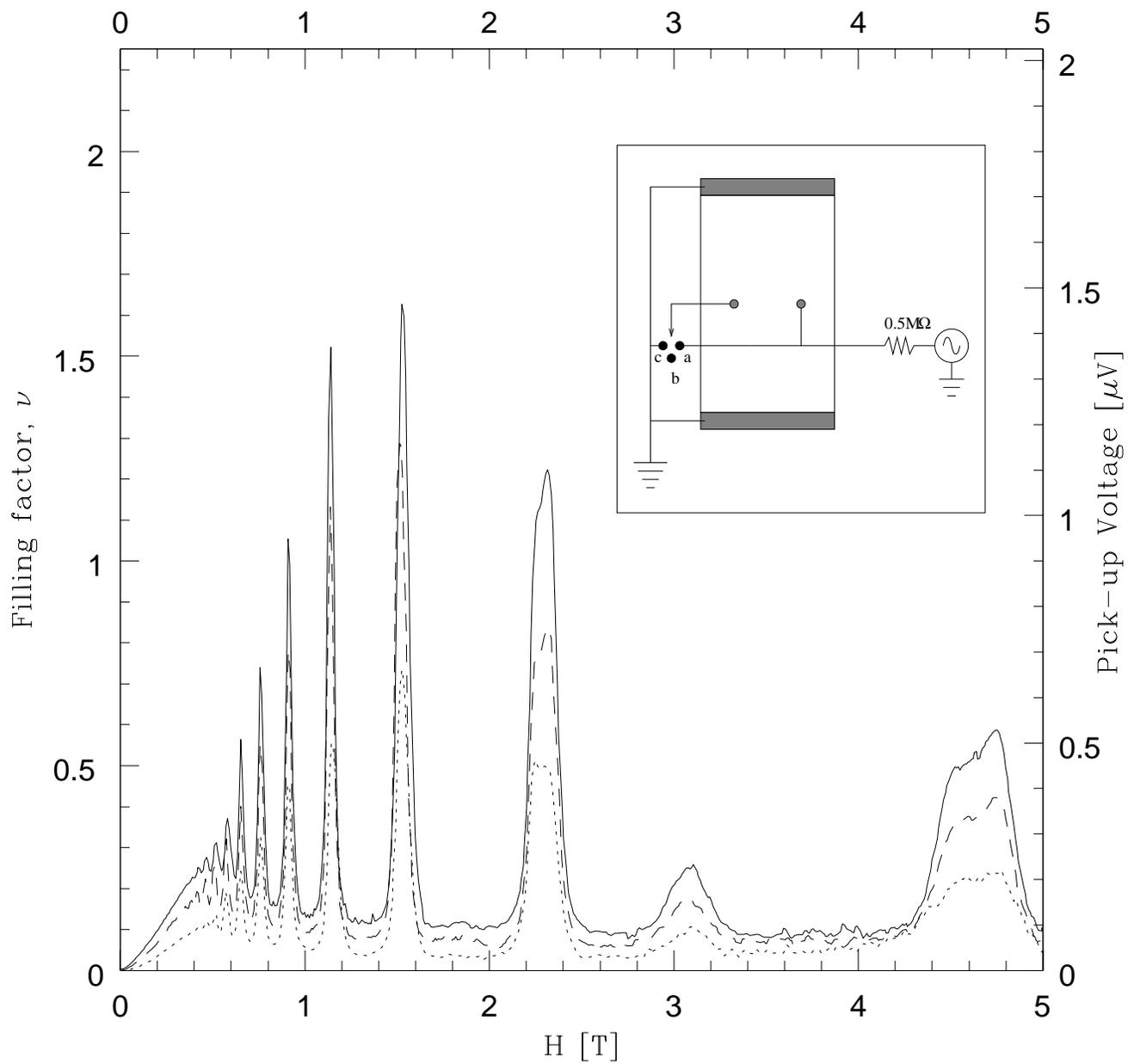
\end{document}